\documentclass[5p]{elsarticle}
\makeatletter
\def\ps@pprintTitle{%
 \let\@oddhead\@empty
 \let\@evenhead\@empty
 \def\@oddfoot{}%
 \let\@evenfoot\@oddfoot}
\makeatother

\usepackage{csquotes}
\usepackage{graphicx,color}
\usepackage{float}
\usepackage[caption=false]{subfig}
\usepackage{amsmath}
\usepackage{amssymb}
\usepackage{multirow}
\usepackage{dsfont}
\usepackage{adjustbox}
\usepackage{pdflscape}
\usepackage{wasysym}

\begin{document}

\title{Wall damage due to oblique high velocity dust impacts}
\author{Panagiotis Tolias$^{a}$, Marco De Angeli$^{b}$, Dario Ripamonti$^{c}$, Svetlana Ratynskaia$^{a}$, Giambattista Daminelli$^{c}$ and Monica De Angeli$^{b}$}
\address{$^a$Space and Plasma Physics - KTH Royal Institute of Technology, Teknikringen 31, 10044 Stockholm, Sweden\\
         $^b$Institute for Plasma Science and Technology, CNR, via Cozzi 53, 20125 Milano, Italy\\
         $^c$Institute of Condensed Matter Chemistry and Energy Technologies, CNR, via Cozzi 53, 20125 Milano, Italy}
\begin{abstract}
\noindent Runaway electron termination on plasma facing components can trigger material explosions that are accompanied by the expulsion of fast solid debris. Due to the large kinetic energies of the ejected dust particles, their subsequent mechanical impacts on the vessel lead to extensive cratering. Earlier experimental studies of high velocity micrometric tungsten dust collisions with tungsten plates focused exclusively on normal impacts. Here, oblique high velocity tungsten-on-tungsten mechanical impacts are reproduced in a controlled manner by a two-stage light gas gun shooting system. The strong dependence of the crater characteristics and crater morphology on the incident angle is documented. A reliable empirical damage law is extracted for the dependence of the crater depth on the incident angle.
\end{abstract}
\maketitle

\section{Introduction}

\noindent The long preservation of the integrity of plasma facing components (PFCs) constitutes a major challenge for future fusion reactors\,\cite{introd01,introd02}. The PFC lifetime is primarily threatened during unplanned transients rather than the quiescent flat-top phase. The potential of intense surface heat loads, that are generated by edge-localized modes, vertical displacement events or major disruptions, to cause considerable wall melting and deformation has been extensively studied\,\cite{introd03} and reliable cost-effective numerical tools with predictive power have already been developed\,\cite{introd04,introd05,introd06}. The potential of extreme volumetric heat loads, that are generated by runaway electron (RE) beams, to cause extended vessel damage has been known for decades\,\cite{introd07}, but it has been investigated far less systematically\,\cite{introd08}. In particular, it has been documented that the RE incidence can trigger PFC explosions that are driven by thermal expansion and/or internal boiling and are accompanied by the expulsion of fast solid micrometric debris\,\cite{introd09,introd10,introd11,introdsu}. The local power handling capabilities at the explosion location are naturally compromised raising the risk of disastrous loss-of-coolant accidents\,\cite{introd12}. Moreover, subsequent impacts of the ejected dust have been observed to lead to extensive cratering spread all over the wall\,\cite{introd09,introdsu}. The typical debris speeds and sizes are of the order of $1\,$km/s and $50\,\mu$m, respectively\,\cite{introd08,introd09,introd10}. Thus, the dust-wall mechanical impacts belong to the high velocity impact regime, $200\lesssim{v}_{\mathrm{imp}}[\mathrm{m/s}]\lesssim4000$, that is characterized by strong deformation of dust and shallow crater formation\,\cite{impactR1,impactR2}.

Unsurprisingly, the indirect non-localized aspect of RE-induced PFC damage, caused by the mechanical impacts of the ejected debris onto PFCs within the line of sight of the RE termination sites, has received much less attention than the direct localized aspect of RE-induced PFC damage. Nevertheless, there are numerous reasons that render the study of high speed impacts important. \emph{First}, this type of damage cannot be constrained to sacrificial limiters or replaceable divertor plates and thus poses a threat to sensitive vessel constituents including diagnostics. Therefore, it is important to quantify the excavated volume of impact craters and assess its consequences. \emph{Second}, for given projectile-target composition, there is a direct correlation between the dust size and impact velocity with the crater depth and diameter\,\cite{impactR3}. Therefore, provided that reliable damage laws are available, contemporary tokamak experiments dedicated to RE-induced PFC damage can utilize witness plates as an independent diagnostic of the ejected dust speed and dust size distributions. These debris properties constitute crucial benchmarks for numerical tools that model RE-driven fragmentation. \emph{Third}, high velocity dust impacts are accompanied by high strain rates, localized heating and projectile \& target fragmentation. Thus, they share many common physics with RE-driven explosions and can be utilized to inform the thermomechanical constitutive laws as well as to enhance the predictive power of tools that simulate RE-driven explosions\,\cite{introd08}.

Recent theoretical investigations dedicated to tungsten-on-tungsten (W-on-W) high velocity impacts were mostly based on classical molecular dynamic (MD) simulations with model interatomic potentials\,\cite{modelli1,modelli2,modelli3,modelli4,modelli5,modelli6}. Due to the inherent computational limitations of such brute-force technniques, these studies either concerned cluster projectiles\,\cite{modelli1,modelli2,modelli3} or nanodust projectiles\,\cite{modelli4,modelli5,modelli6}. Thus, although such MD investigations provide invaluable microscopic insights on the impact microphysics\,\cite{modelli4,modelli5,modelli6}, the observed crater scalings require size extrapolations of many orders of magnitude to reach the dust and crater dimensions that are relevant to tokamaks. To our knowledge, other computational methods which can adequately describe high velocity impacts, such as smoothed particle hydrodynamics\,\cite{SPHtref1,SPHtref2} and peridynamic theory\,\cite{peridyn1,peridyn2}, have not yet been applied to the W-on-W case.

Recent experimental investigations dedicated to W-on-W high velocity impacts were based on aerodynamic dust acceleration achieved with the two-stage light gas guns\,\cite{lightga1,lightga2}. Normal incidence room temperature high velocity W-on-W impacts were exhaustively studied, ultimately leading to the extraction of reliable empirical damage laws that express the crater depth and diameter as analytical functions of the dust size and dust speed\,\cite{normaHVW}. For normal incidence and room temperature dust, the dependence on the target temperature was also investigated in experiments with resistively heated and liquid nitrogen cooled bulk samples\,\cite{tempeHVW}. The dependence on the incidence angle has not been yet investigated for W-on-W impacts. However, general theoretical arguments and experiments with other materials have demonstrated that the incident angle dependence is significant and concerns not only the crater dimensions but also the crater morphology\,\cite{obliqgen,obliqBur,obliqGla,obliqice}.

The present experimental investigation provides the missing data on the incident angle dependence of high-velocity impacts of micrometric W dust on bulk W targets. Highly spherical nearly monodisperse $63\,\mu$m diameter W dust is prepared that is accelerated by means of a two-stage light gas gun towards a tilted bulk W target. The dust impact speeds vary controllably within $\sim2000-3000\,$m/s, while the incident angles vary within $0^{\circ}-80^{\circ}$ with respect to the surface normal. The crater characteristics (depth, length, width) are then obtained by means of a scanning electron microscope, an optical microscope and a mechanical profiler. A simple scaling law is extracted for the dependence of the crater depth on the incident angle, while the more elaborate dependence of the crater length \& width on the incident angle is also documented. The results are discussed in view of MD studies for W-on-W\,\cite{modelli6} and experimental studies for other metals\,\cite{obliqBur}.

\section{Theoretical aspects}\label{sec:theory}

In the case of micrometric spherical projectiles normally impinging on bulk semi-infinite planar targets (both composed of refractory metals), four impact speed ranges can be distinguished. It is pointed out that the following speed thresholds are indicative due to their dependence on the projectile size and the material composition. \textbf{(1)} In the low velocity range, $v_{\mathrm{imp}}\lesssim5\,$m/s, the projectiles do not rebound but they stick to the target, since the irreversible adhesive work suffices to dissipate the entire incident kinetic energy\,\cite{regime01,regime02,regime03}. Plastic dissipation might also play a role, but only towards the upper velocity threshold. \textbf{(2)} In the moderate velocity range, $5\,\mathrm{m/s}\lesssim{v}_{\mathrm{imp}}\lesssim200\,$m/s, the projectiles inelastically rebound from the target, but can be assumed to retain their morphology. The kinetic energy difference is expended on adhesive work, viscoelastic dissipation as well as plastic work\,\cite{regime03,regime04,regime05,regime06,regime07}. \textbf{(3)} In the high velocity range, $200\,\mathrm{m/s}\lesssim{v}_{\mathrm{imp}}\lesssim4000\,$m/s, the projectiles are severely flattened due to the strong plastic deformation or partially fragmented when ultimate strength limits are exceeded. These impacts are also characterized by shallow cratering and near-surface melting on the target\,\cite{regime08,regime09,regime10,regime11}. \textbf{(4)} Finally, in the hyper velocity range, ${v}_{\mathrm{imp}}\gtrsim4000\,$m/s, that is dominated by shockwave physics, enormous pressures and temperatures are locally generated so that both projectiles and targets extensively melt and vaporize\,\cite{regime12,regime13,regime14,regime15,regime16}. These impacts are also characterized by deep target cratering, fast secondary ejecta release and even plasma formation.

The high velocity range of interest can be subdivided into four regimes. \textbf{(3a)} In the plastic deformation regime, $200\,\mathrm{m/s}\lesssim{v}_{\mathrm{imp}}\lesssim500\,$m/s, plastic work dominates the energy exchange, but fracture limits are not exceeded. Consequently, the projectiles become severely flattened, shallow craters are generated on the target and the projectile rebound speeds are very low\,\cite{regime08,regime09,regime17,regime18}. \textbf{(3b)} In the impact bonding regime, $500\,\mathrm{m/s}\lesssim{v}_{\mathrm{imp}}\lesssim1000\,$m/s, there is augmented local plastic deformation and elevated temperatures at the interface. The projectile sticks to the target mainly due to metallurgical
bonding and mechanical interlocking\,\cite{regime10,regime11,regime19}. \textbf{(3c)} In the partial-disintegration partial-sticking regime, $1000\,\mathrm{m/s}\lesssim{v}_{\mathrm{imp}}\lesssim2500\,$m/s, there is partial projectile fragmentation and pronounced target crater formation with projectile fragments sticking inside the crater\,\cite{tempeHVW}. \textbf{(3d)} In the partial-disintegration no-sticking regime, $2500\,\mathrm{m/s}\lesssim{v}_{\mathrm{imp}}\lesssim4000\,$m/s, there is also partial projectile fragmentation and pronounced target crater formation but without projectile fragments remaining stuck inside the crater\,\cite{tempeHVW}.

It is also emphasized that the above speed thresholds refer to normal impacts, but there should also be a strong dependence on the impact angle of the projectile. Note that it has not even been established that all impact regimes of the high velocity range can be realized for very oblique impacts. In this investigation, we shall focus on projectile total speeds that belong to both partial disintegration regimes (3c+3d), since these regimes are the most harmful for the target and since these regimes are known to lead to simple correlations for the crater characteristics versus the impact speed.

\section{Experimental aspects}\label{sec:experimental}

The W dust particles are loaded inside the cavity of a macro-scale projectile (sabot). The sabot is accelerated to the targeted speeds by means of a two-stage light gas gun, see Refs.\cite{normaHVW,tempeHVW} for technical details. The dust particles are separated from the sabot at the end of the launch tube and free stream through a vacuum chamber towards the W target. The oblique impacts are simply realized by tilting the target with respect to the line-of-sight direction.

A nearly monodisperse W dust sub-population with diameters centered around $63\,\mu$m was meshed out using a sequence of sieves from a polydisperse batch with a nominal size distribution within $45-90\,\mu$m. Such a mean dust size is comparable to the most probably sizes of fast solid debris that emanate from the RE-PFC interaction region in contemporary tokamaks\,\cite{introd09,introd10}. The original batch was purchased from \enquote{TEKNA Plasma Systems} and was characterized by high-sphericity, good electrical conductivity and low internal porosity. These favorable dust properties are enforced by a spheroidization and compactification technique that is based on precursor powder melting by a radio frequency inductively coupled plasma torch followed by free-fall resolidification. TEKNA W dust has been successfully utilized in dust-in-tokamak studies concerning injection\,\cite{TEKNAex1}, accumulation\,\cite{TEKNAex2}, remobilization\,\cite{TEKNAex3,TEKNAex4}, adhesion\,\cite{TEKNAex5,TEKNAex6} and mechanical impacts\,\cite{normaHVW,tempeHVW}.

The polished W targets can be safely considered to be perfectly planar, perfectly smooth and semi-infinite as far as the realized high velocity impacts are concerned. The target radius of curvature is orders of magnitude larger than the dust diameter, while the dust diameter is orders of magnitude larger than the surface roughness length scale. Furthermore, the W targets have a $1\,$mm thickness, which is nearly $200\times$ larger than the deeper crater realized in our impact experiments. Thus, finite thickness effects are not expected to play a role regardless of the impact speed

\begin{figure}
\centering
\includegraphics[width = 3.5in]{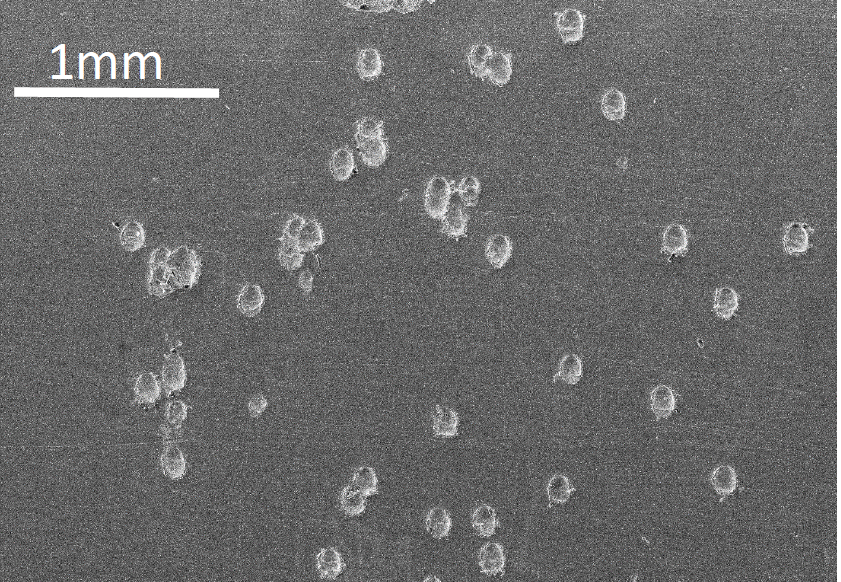}
\caption{SEM image of a damaged bulk W target after the high velocity impact of spherical $63\,\mu$m W dust (here with $1988\,$m/s and $60^{\circ}$). Nearly half of the realized impact craters are overlapping. Only the isolated craters are considered in the statistical analysis of the crater morphology.}\label{fig:overlappingSEM}
\end{figure}

Three nominal impact speeds were targeted that all belong to the two partial disintegration regimes: $2000\,$m/s (but in practice $1988-2052\,$m/s), $2500\,$m/s (but in practice $2412-2556\,$m/s) and $3000\,$m/s (but in practice $2921-3108\,$m/). Six impact angles were realized that range from normal ($0^{\circ}$) to grazing ($80^{\circ}$). Overall, the above translate to $18$ impact tests. Despite the fact that the light gas gun acceleration technique leads to a large number of W-on-W impacts, the useful crater statistics are rather restricted. This is a consequence of the unavoidable overlapping of the craters (which are easily identifiable) that need to be excluded from the statistical analysis. A characteristic example featuring both overlapping and isolated craters is provided in Fig.\ref{fig:overlappingSEM}.

\begin{figure}
\centering
\includegraphics[width = 3.5in]{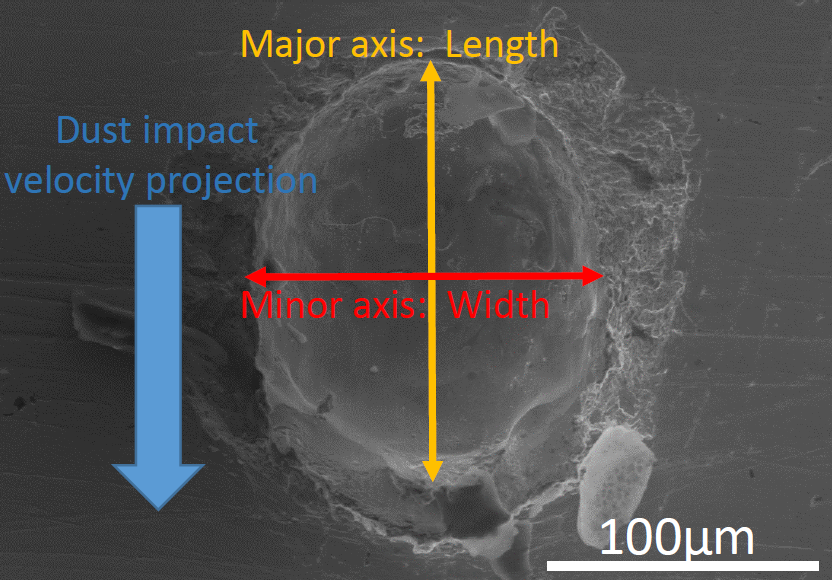}
\caption{SEM image of damaged bulk W targets after the high velocity impact of spherical $63\,\mu$m W dust (here with $2451\,$m/s and $45^{\circ}$). The definition of the 2D crater figures of merit (length and width) within the elliptical approximation. The major axis is parallel to the projection of the dust impact velocity on the target surface.}\label{fig:morphologySEM}
\end{figure}

\begin{figure*}
\centering
\includegraphics[width = 7.2in]{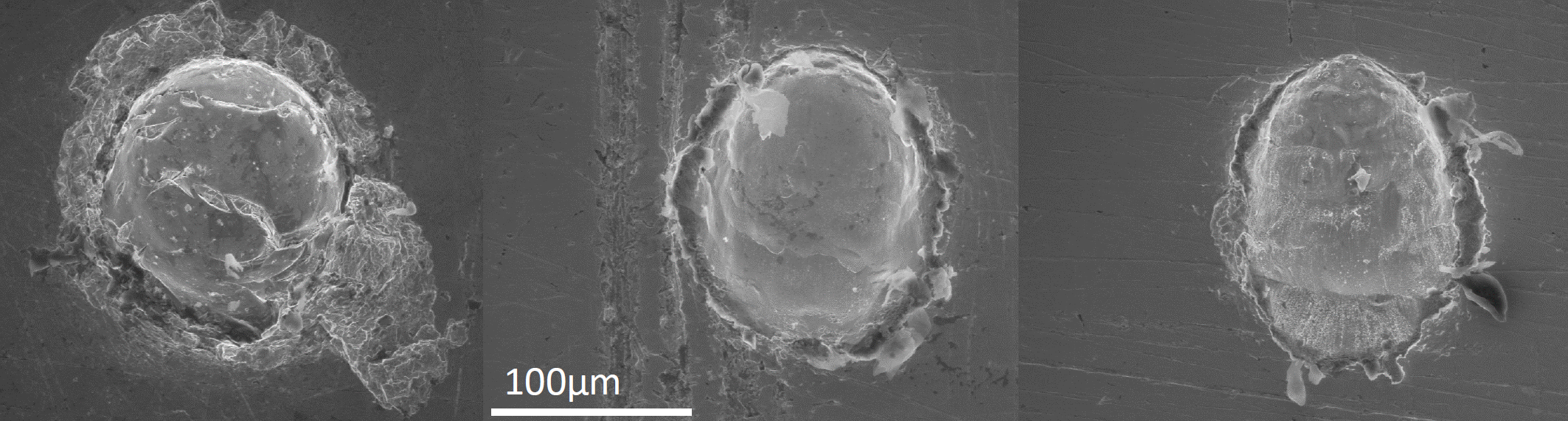}
\caption{SEM images of damaged bulk W targets after the high velocity impact of spherical $63\,\mu$m W dust. Here for $2052\,$m/s and $30^{\circ}$ (left), $2036\,$m/s and $45^{\circ}$ (center), $1988\,$m/s and $60^{\circ}$ (right). For a near-constant impact speed and dust size, the crater morphology transitions from near-circular to near-elliptical as the impact angle increases.}\label{fig:morphologytransition}
\end{figure*}

\begin{figure}
\centering
\includegraphics[width = 3.5in]{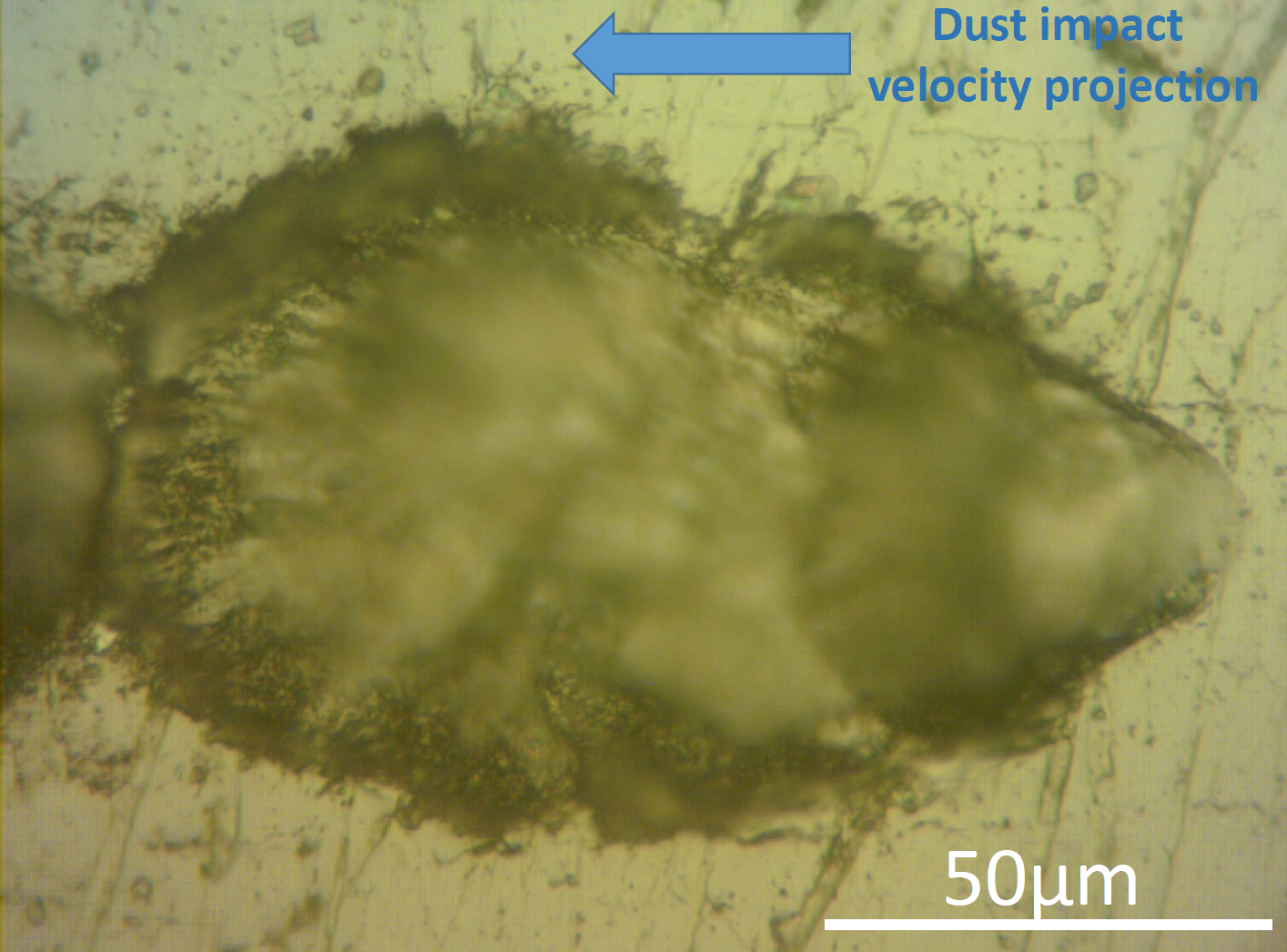}
\caption{Optical image of a damaged bulk W target after the high velocity impact of spherical $63\,\mu$m W dust (here with $2412\,$m/s and $80^{\circ}$). At grazing angles, clear non-elliptical features emerge. The elliptical approximation is still applicable, but becomes less accurate. The shallow head feature is barely visible at the left. The projection of the dust impact velocity on the target surface is also illustrated.}\label{fig:nonelliptical}
\end{figure}

\begin{figure}
\centering
\includegraphics[width = 3.5in]{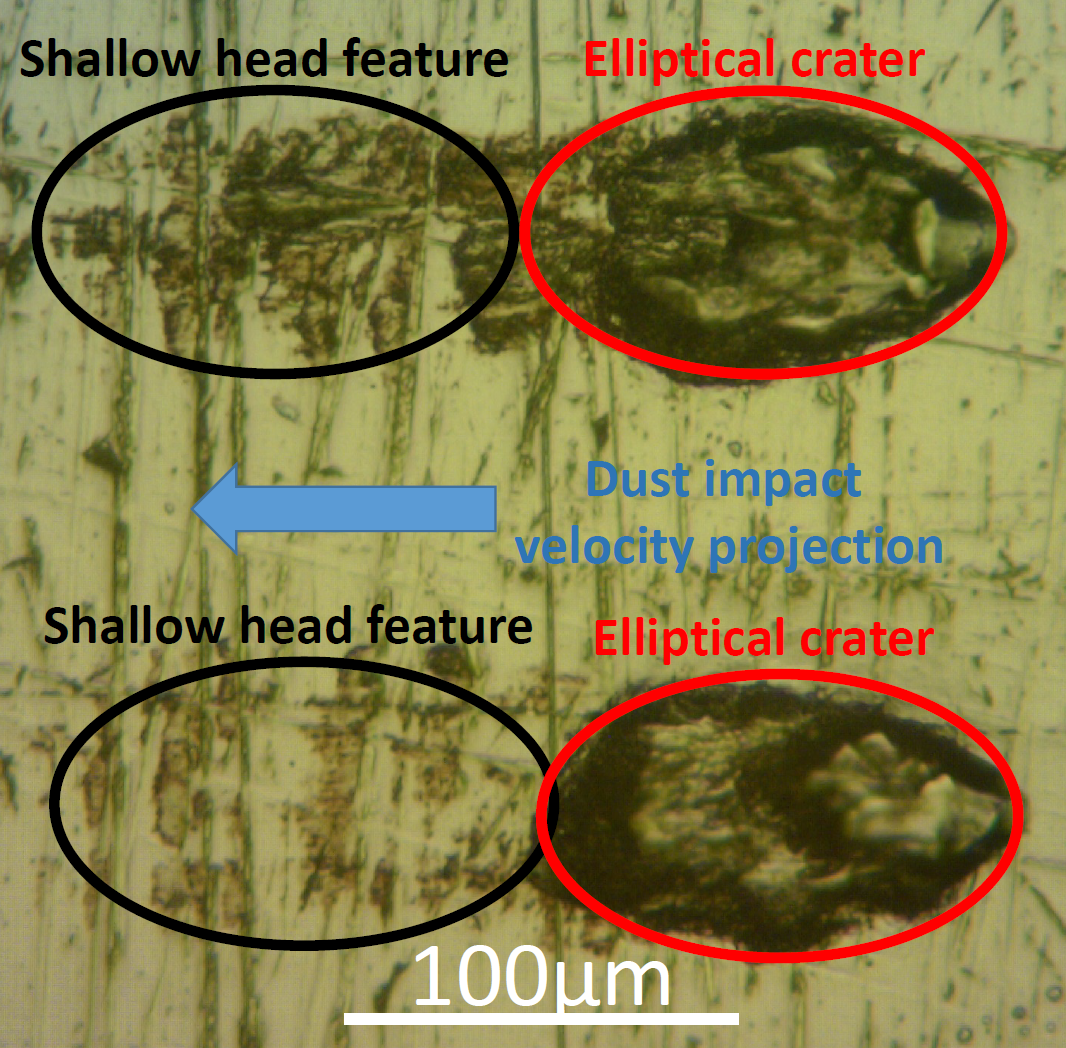}
\caption{Optical image of a damaged bulk W target after the high velocity impact of spherical $63\,\mu$m W dust (here with $2008\,$m/s and $80^{\circ}$). Two neighboring elliptical craters that feature a shallow head. The projection of the dust impact velocity on the target surface is also illustrated.}\label{fig:obliquehead}
\end{figure}

As the impact angle becomes more oblique, the crater morphology switches from near-circular to near-elliptical with the major axis along the target projection of the impact velocity. Thus, within a half ellipsoid assumption, the crater morphology can be adequately quantified with three figures of merit: the crater depth, length (major axis) and width (minor axis), see Fig.\ref{fig:morphologySEM}. The circular-to-elliptical transition of the morphology can be seen in Fig.\ref{fig:morphologytransition}. As the dust impact angle becomes grazing ($70^{\circ},80^{\circ}$), the crater morphology exhibits small deviations from the pure elliptical shape, see Fig.\ref{fig:nonelliptical} for a characteristic example. These \enquote{fish-like} deviations can be quantified by defining two values for the minor diameter, but they have been judged to be unimportant in view of the omnipresent experimental uncertainties. Moreover, at grazing incidence ($70^{\circ},80^{\circ}$), the craters often feature an elongated head which is not accounted for in the length estimation being very shallow, as clearly discerned from Fig.\ref{fig:obliquehead}. It is speculated that the head feature is caused by the secondary impacts of high velocity fragments generated from the main dust impact, which also justifies the exclusion of these features from the morphological evaluation. This speculation is based on the fact that, for a constant oblique angle, the head-to-crater distance seems to increase with the impact speed.

Each of the $18$ W targets has been mapped by means of a Scanning Electron Microscope (SEM) and a precision optical microscope. The crater depth has been evaluated with the optical microscope. Owing to inherent difficulties in distinguishing the crater boundary in 2D SEM images, the crater length and crater width have been evaluated on the basis of the SEM images for impact angles less than $60^{\circ}$ and on the basis of the optical images for impact angles larger than $60^{\circ}$.

Finally, let us discuss the experimental uncertainties: \textbf{(1)} The dust diameter uncertainties are found to be $\pm3\,\mu$m from the spread of the nearly monodisperse dust size distribution. \textbf{(2)} The dust impact speed uncertainties are generally less than $\pm1\%$. The speed is measured through the dust transit time between two laser sheets with the error stemming from the relatively large thickness of the individual beams ($1\,$mm) when compared to the beam spacing ($100\,$mm)\,\cite{normaHVW,tempeHVW}. Essentially, the dust cloud speed is measured which should be identical to the dust particle speed, given the near-constant small cloud width recorded by the laser sheets\,\cite{normaHVW}. \textbf{(3)} The impact angle uncertainties are generally less than $0.5^{\circ}$; the W plate can be considered as perfectly planar, the plate holder inclination angle is accurate within half degree and the acceleration mechanism is purely normal with respect to the launch tube axis. \textbf{(4)} The target crater dimension uncertainties are of statistical and instrumental origin. The former uncertainties are featured in Table \ref{tab:statistics_oblique}. Concerning the latter uncertainties: the crater depth is measured with a precision optical microscope of $0.5\,\mu$m sensitivity with a $\pm3\,\mu$m associated uncertainty\,\cite{normaHVW,tempeHVW}, while the length and width are evaluated on the basis of SEM and optical images with a $10-20\%$ associated uncertainty depending on the incident angle.

\begin{table}
  \centering
  \caption{W-on-W oblique impacts realized by the two-stage light gas gun. The mean dust diameter is $63\,\mu$m, the dust is near-perfectly spherical and the target is near-perfectly planar. In what follows; $v_{\mathrm{imp}}$ denotes the dust impact speed, $\#$ the number of craters, $\theta$ the nominal impact angle with respect to the surface normal, $H$ the average crater depth, $L$ the average crater length (major axis of the ellipse) and $W$ the average crater width (minor axis of the ellipse). The statistical errors are also included for $H,L,W$.}\label{tab:statistics_oblique}
\begin{tabular}{c c c c c c c c c}
$v_{\mathrm{imp}}$ & $\#$  & $\theta$     &  $H$          & $L$          & $W$       \\
(m/s)              &       & $(^{\circ})$ &  ($\mu$m)     & ($\mu$m)     & ($\mu$m)  \\ \hline\hline
$2039$             & 78    & 0            &  $43\pm7$     & $119\pm7$    & $119\pm7$ \\
$2052$             & 45    & 30           &  $31\pm4$     & $135\pm8$    & $135\pm8$ \\
$2036$             & 17    & 45           &  $34\pm3$     & $147\pm6$    & $118\pm7$ \\
$1988$             & 34    & 60           &  $23\pm3$     & $152\pm9$    & $110\pm6$ \\
$2001$             & 29    & 70           &  $16\pm3$     & $126\pm11$   & $81\pm6$  \\
$2008$             & 29    & 80           &  $6\pm1$      & $85\pm13$    & $51\pm6$  \\ \hline
$2485$             & 37    & 0            &  $51\pm8$     & $129\pm7$    & $129\pm7$ \\
$2556$             & 51    & 30           &  $45\pm6$     & $145\pm8$    & $145\pm8$ \\
$2451$             & 44    & 45           &  $44\pm3$     & $156\pm11$   & $124\pm9$ \\
$2465$             & 13    & 60           &  $30\pm3$     & $161\pm9$    & $117\pm8$ \\
$2556$             & 23    & 70           &  $16\pm3$     & $149\pm15$   & $86\pm10$ \\
$2412$             & 30    & 80           &  $7\pm1$      & $116\pm14$   & $60\pm5$  \\ \hline
$3108$             & 33    & 0            &  $62\pm11$    & $142\pm21$   & $142\pm21$\\
$2933$             & 13    & 30           &  $57\pm4$     & $160\pm12$   & $160\pm12$\\
$2950$             & 12    & 45           &  $50\pm4$     & $170\pm12$   & $170\pm12$\\
$2927$             & 28    & 60           &  $36\pm2$     & $176\pm11$   & $134\pm9$ \\
$2921$             & 15    & 70           &  $22\pm2$     & $167\pm8$    & $115\pm6$ \\
$2934$             & 20    & 80           &  $9\pm1$      & $130\pm14$   & $65\pm5$  \\ \hline\hline
\end{tabular}
\end{table}

\section{Analysis}\label{sec:anal}

\noindent Our earlier comprehensive light gas gun experiments that focused on W-on-W cratering due to \emph{normal} high velocity impacts of micron-sized dust, established empirical W-on-W damage laws which describe the dependence of the crater depth and crater diameter on the dust size and dust impact speed\,\cite{normaHVW}. The empirical correlations read as
\begin{align}
H_{\mathrm{n}}(D_{\mathrm{d}},v_{\mathrm{imp}})&=0.0000114(D_{\mathrm{d}})^{1.264}(v_{\mathrm{imp}})^{1.282}\,,\label{eq:normaldepth}\\
L_{\mathrm{n}}(D_{\mathrm{d}},v_{\mathrm{imp}})&=0.0330(D_{\mathrm{d}})^{1.005}(v_{\mathrm{imp}})^{0.527}\label{eq:normallength}\,,
\end{align}
with the crater depth $H_{\mathrm{n}}$, crater length $L_{\mathrm{n}}$ and dust diameter $D_{\mathrm{d}}$ measured in $\mu$m, while the normal impact speed $v_{\mathrm{imp}}$ is measured in m/s. These damage laws are based on experiments with $D_{\mathrm{d}}=51,63,76\,\mu$m sized spherical W dust that has been aerodynamically accelerated to $v_{\mathrm{imp}}=1500-3500\,$m/s normal impact speeds.

Let us return to our new light gas gun experiments that focus on W-on-W cratering due to \emph{oblique} high velocity impacts of $63\,\mu$m dust. The crater dimensions (depth, length, width) and their standard deviations (only the statistical errors without including the instrumental errors) are listed in Table \ref{tab:statistics_oblique}. Based on Eqs.(\ref{eq:normaldepth},\ref{eq:normallength}), it can be inferred that the statistical errors in the crater dimensions can be mostly attributed to the dust size distribution. For instance, noting that $D_{\mathrm{d}}=63\pm3\mu$m, we have $H_{\mathrm{n}}(66,2500)-H_{\mathrm{n}}(60,2500)\simeq6\,\mu$m and $L_{\mathrm{n}}(66,2500)-L_{\mathrm{n}}(60,2500)\simeq12\,\mu$m. In addition, based on Eqs.(\ref{eq:normaldepth},\ref{eq:normallength}), it can be deduced that the minor variations around the targeted dust impact speed barely affect the crater dimensions. For instance, we have $H_{\mathrm{n}}(63,2052)-H_{\mathrm{n}}(63,1988)\sim1.5\,\mu$m and $L_{\mathrm{n}}(63,2052)-L_{\mathrm{n}}(63,1988)\sim2\,\mu$m for the lowest targeted impact speed of $2000\,$m/s.

\begin{figure*}
\centering
\includegraphics[width = 6.9in]{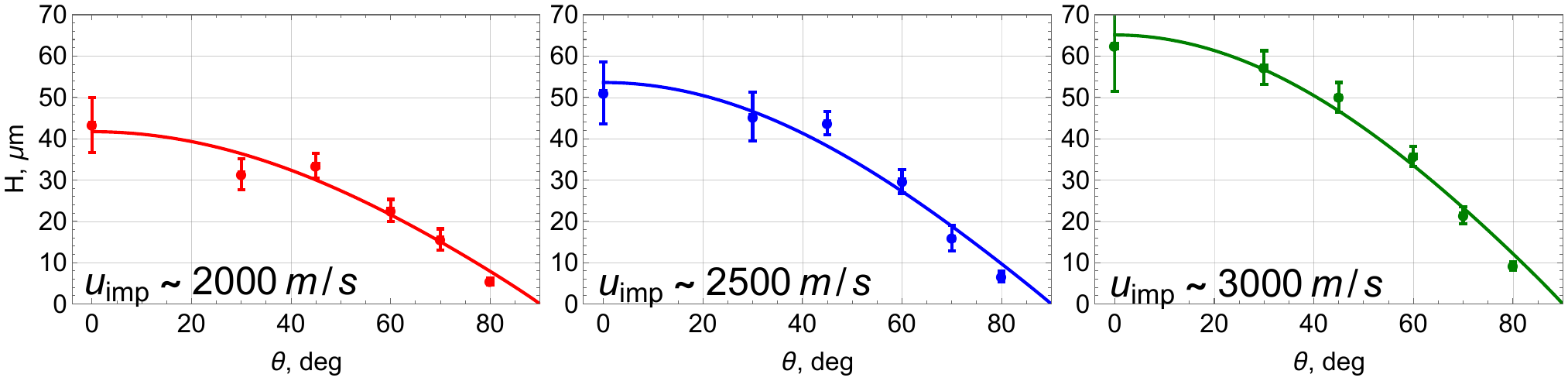}
\caption{The crater depth as a function of the incident angle with respect to the surface normal for $63\mu$m spherical W dust impinging on bulk planar W plates with an impact speed of (a) $v_{\mathrm{imp}}\sim2000$m/s, (b) $v_{\mathrm{imp}}\sim2500$m/s, (c) $v_{\mathrm{imp}}\sim3000$m/s. The data points and the statistical errors are listed in the fourth column of Table \ref{tab:statistics_oblique}. The solid lines correspond to the $a(\cos{\theta})^b$ least square fits.}\label{fig:crater_depth}
\end{figure*}

Within the combined statistical and measurement uncertainties, for the three probed impact speeds, the \emph{crater depth} is discerned to be a monotonically decreasing function of the incident angle with respect to the normal. In particular, regardless of the impact speed, its incident angle dependence can be accurately fitted to a simple power law $H(\theta)/H_{\mathrm{n}}=a(\cos{\theta})^b$. The pre-factors are found to be $a=0.963,1.049,1.043$, while the exponents are given by $b=0.949,0.975,0.958$ for $v_{\mathrm{imp}}\sim2000,2500,3000$\,m/s, respectively. The experimental data and the least square fits are plotted in Fig.\ref{fig:crater_depth}. The fitting parameters are essentially independent of the impact speed and can be approximated by $a\simeq1$, $b\simeq1$, which leads to the empirical damage law
\begin{align}
\frac{H(D_{\mathrm{d}},v_{\mathrm{imp}},\theta)}{H_{\mathrm{n}}(D_{\mathrm{d}},v_{\mathrm{imp}})}=\cos{\theta}\,.
\end{align}
This linear cosine dependence might hold quite generally for projectiles and targets of the same composition, since it has also been observed in the experiments of Burchell \& Mackay featuring mm-size aluminum spheres impinging on bulk aluminum with $v_{\mathrm{imp}}\sim5000$\,m/s ($b=0.99$) and mm-size stainless steel spheres impinging on bulk stainless steel with $v_{\mathrm{imp}}\sim5000$\,m/s ($b=0.91$)\,\cite{obliqBur} as well as in the MD simulations of Dwivedi \& Fraile featuring tungsten nanodust impinging on bulk tungsten with $v_{\mathrm{imp}}=3000$\,m/s ($b=1.09$)\,\cite{modelli6}. It is worth pointing out that, for the three probed impact speeds, the crater depth for $\theta=45^{\circ}$ is consistently underestimated by the cosine power law dependence. This might be purely coincidental given the uncertainties, but it might also be connected with the morphological transition from circular to elliptical craters which also occurs around $\theta=45^{\circ}$, see also Fig.\ref{fig:morphologytransition}. More experimental data are required to elucidate this observation.

\begin{figure}
\centering
\includegraphics[width = 3.0in]{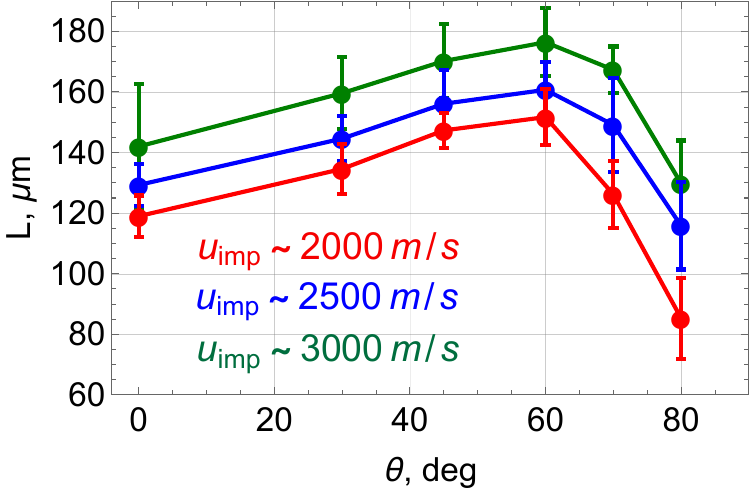}
\caption{The crater length as a function of the incident angle with respect to the surface normal for $63\mu$m spherical W dust impinging on bulk planar W plates with impact speeds of $v_{\mathrm{imp}}\sim2000$m/s (red), $v_{\mathrm{imp}}\sim2500$m/s (blue), $v_{\mathrm{imp}}\sim3000$m/s (dark green). The data points and the statistical errors are listed in the fifth column of Table \ref{tab:statistics_oblique}. The solid lines serve as a guide-to-the-eye.}\label{fig:crater_length}
\end{figure}

On the other hand, the \emph{crater length} is not a monotonically decreasing function of the incident angle $\theta$ and does not have an elementary power-law dependence on $\cos{\theta}$. In particular, as illustrated in Fig.\ref{fig:crater_length}, regardless of the impact speed, there is a maximum of the crater length that is reached at $\theta\sim60^{\circ}$. Despite the statistical errors, the three curves for $v_{\mathrm{imp}}\sim2000,2500,3000\,$m/s are clearly displaced versions of the same master curve. As expected, for a constant incident angle, the crater length is a monotonically increasing function of the impact speed. It is worth pointing out that a similar non-monotonic dependence of the crater length has been observed in the experiments of Burchell \& Mackay featuring mm-size spheres\,\cite{obliqBur}, whereas a weak power-law dependence of the type $L/L_{\mathrm{n}}=(\cos{\theta})^{0.14}$ was inferred from the MD simulations of Dwivedi \& Fraile featuring W nanodust impinging on bulk W\,\cite{modelli6}.

\begin{figure}
\centering
\includegraphics[width = 3.0in]{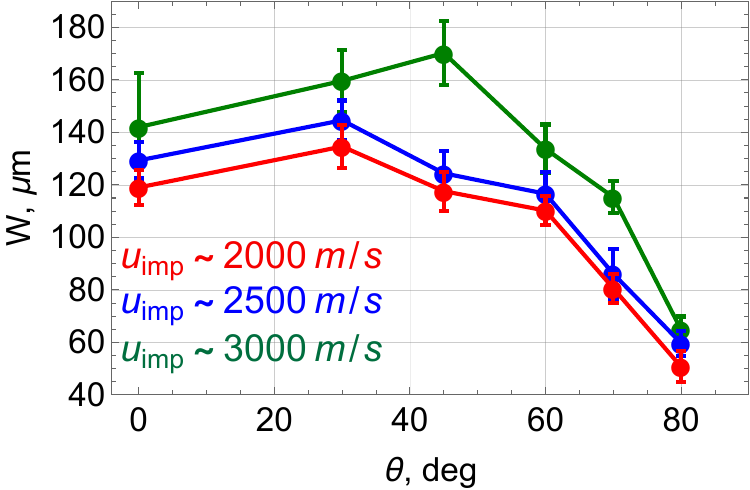}
\caption{The crater width as a function of the incident angle with respect to the surface normal for $63\mu$m spherical W dust impinging on bulk planar W plates with impact speeds of $v_{\mathrm{imp}}\sim2000$m/s (red), $v_{\mathrm{imp}}\sim2500$m/s (blue), $v_{\mathrm{imp}}\sim3000$m/s (dark green). The data points and the statistical errors are listed in the sixth column of Table \ref{tab:statistics_oblique}. The solid lines serve as a guide-to-the-eye.}\label{fig:crater_width}
\end{figure}

Moreover, the \emph{crater width} is also not a monotonically decreasing function of the incident angle $\theta$ and does not have a simple dependence on $\cos{\theta}$. It should be noted that up to $\theta\sim30-45^{\circ}$, the eccentricity of the crater is approximately equal to unity, thus the crater width is taken to be equal to the crater length within the statistical uncertainties. As illustrated in Fig.\ref{fig:crater_width}, regardless of the impact speed, there is a maximum of the crater width that is reached at $\theta\sim30-45^{\circ}$, which is strongly correlated with the switch of the crater morphology from circular to elliptical. Note that the maxima of the crater width and the crater length do not appear at the same incident angle. Despite the statistical errors, the two curves for $v_{\mathrm{imp}}\sim2000,2500\,$m/s are clearly displaced versions of the same master curve; this does not apply for $v_{\mathrm{imp}}\sim3000\,$m/s due to the $\theta=45^{\circ}$ data point. As expected, for a constant incident angle, the crater width is also a monotonically increasing function of the impact speed. It is again worth pointing out that a similar non-monotonic dependence of the crater width has been observed in the experiments of Burchell \& Mackay featuring mm-size spheres\,\cite{obliqBur}, whereas a power-law dependence of the type $W/W_{\mathrm{n}}=(\cos{\theta})^{0.44}$ was inferred from the MD simulations of Dwivedi \& Fraile featuring tungsten nanodust impinging on bulk tungsten\,\cite{modelli6}.

\begin{figure}
\centering
\includegraphics[width = 3.0in]{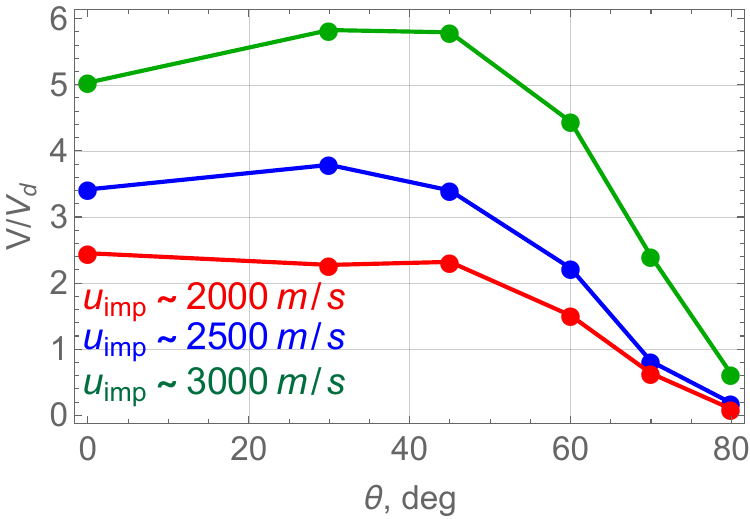}
\caption{The crater volume normalized by the dust volume as a function of the incident angle with respect to the surface normal for $63\mu$m spherical W dust impinging on bulk planar W plates with impact speeds of $v_{\mathrm{imp}}\sim2000$m/s (red), $v_{\mathrm{imp}}\sim2500$m/s (blue), $v_{\mathrm{imp}}\sim3000$m/s (dark green). The volume is computed within a half-ellipsoid assumption with the data of Table \ref{tab:statistics_oblique}. The solid lines serve as a guide-to-the-eye.}\label{fig:crater_volume}
\end{figure}

Finally, the \emph{crater volume} is evaluated for all the impact speed and incident angle combinations under the oversimplifying assumption that the crater is a half-ellipsoid. In Fig.\ref{fig:crater_volume}, we plot the ratio of the crater half-ellipsoid volume $V=(1/6)\pi{H}LW$ over the dust spherical volume $V_{\mathrm{d}}=(1/6)\pi{D_{\mathrm{d}}}^3$. The propagation of the statistical errors has not been followed, since it is not possible to quantify the additional error due to the half-ellipsoid assumption. It is again observed that the crater volume is not a monotonically decreasing function of the incident angle $\theta$ and does not have a simple dependence on $\cos{\theta}$. In particular, there is a relatively shallow maximum of the crater volume that is reached at $\theta\sim30-45^{\circ}$. It is also worth noting that the excavated volume is larger than the dust volume, unless the incident angle is grazing $\theta\sim80^{\circ}$ and that the maximum excavated volume is roughly six times the dust volume. We point out that the MD simulations of Dwivedi \& Fraile featuring tungsten nanodust impinging on bulk tungsten yield a power-law dependence of the type $V/V_{\mathrm{n}}=(\cos{\theta})^{2.5}$\,\cite{modelli6}, which is not confirmed by our experiments.

\section{Summary and future work}

\noindent The first experimental study of the oblique impact of high velocity spherical W dust on bulk W targets has been carried out focusing on the partial disintegration regime. The controlled dust impacts were realized by means of a two-stage light gas gun and the post-mortem surface analysis of the target led to the documentation of the strong dependence of the crater morphology and the crater characteristics on the dust impact angle. This culminated in the extraction of a reliable empirical damage law for the monotonic dependence of the crater depth on the impact angle. It is worth noting that, in agreement with experimental results for different metals but in contrast to MD simulation results for W-on-W, the dependence of the crater length and the crater width on the impact angle was observed to be non-monotonic.

The results are highly relevant for the indirect component of the wall damage induced by explosion-like runaway electron dissipation, which is caused by the high velocity impacts of the ejected solid debris with the surrounding vessel. Given the explosive nature of the runaway electron incidence, these impacts are more likely to be oblique than normal, thus wall cratering predictions should be based on empirical damage laws that include the strong dependence on the impact angle.

In general, tungsten wall cratering due to high velocity spherical tungsten dust impacts depends on the dust speed, the dust impact angle, the target temperature and the dust temperature. With the former dependencies elucidated, future work should focus on the effect of elevated dust temperatures. In particular, the temperatures of the ejected fast solid debris can be expected to span the range from near-room temperature up to the melting point. The simultaneous controlled heating and acceleration of micrometric spherical dust poses an experimental challenge. In the case of light gas guns, heating can be achieved after the acceleration stage by substituting the vacuum chamber with a gas chamber (thus mimicking atmospheric re-entry heating mechanisms).

\section*{Acknowledgments}

\noindent The work has been performed within the framework of the EUROfusion Consortium,\,funded by the European Union via the Euratom Research and Training Programme (Grant Agreement No\,101052200 - EUROfusion). Views and opinions expressed are however those of the authors only and do not necessarily reflect those of the European Union or European Commission.\,Neither the European Union nor the European Commission can be held responsible for them.

\end{document}